\documentclass[preprint2]{emulateapj-rtx4}

\slugcomment{To appear in ApJL}

\usepackage{txfonts} \usepackage{epsfig} \usepackage{graphicx}
\usepackage{natbib} 

\begin{document}

  \title{ALMA 690 GHz observations of IRAS 16293-2422B: Infall in a highly optically-thick disk}

\shortauthors{Zapata, et al.}

\author{Luis A. Zapata\altaffilmark{1}, Laurent Loinard\altaffilmark{1,2}, 
Luis F. Rodr\'\i guez\altaffilmark{1,3}, Vicente Hern\'andez\altaffilmark{1}, \\
Satoko Takahashi\altaffilmark{4}, Alfonso Trejo\altaffilmark{4}, and 
B\'ereng\`ere Parise\altaffilmark{2}} 

\altaffiltext{1}{Centro de Radioastronom\'\i a y Astrof\'\i sica, UNAM, Apdo. Postal 3-72
  (Xangari), 58089 Morelia, Michoac\'an, M\'exico}
\altaffiltext{2}{Max-Planck-Institut f\"{u}r Radioastronomie, Auf dem H\"ugel
  69, 53121, Bonn, Germany} 
\altaffiltext{3}{Astronomy Department, Faculty of Science, King Abdulaziz University,
P.O. Box 80203, Jeddah 21589, Saudi Arabia}
\altaffiltext{4}{Academia Sinica Institute of Astronomy and Astrophysics, P.O. Box 23-141, 
Taipei 10617, Taiwan}
\begin{abstract} 
We present sensitive, high angular resolution ($\sim$ 0.2 arcsec)
submillimeter continuum and line observations of IRAS 16293-2422B made
with the Atacama Large Millimeter/Submillimeter Array (ALMA). 
The 0.45 mm continuum observations reveal a single and very compact 
source associated with IRAS 16293-2422B.  This submillimeter source has a 
deconvolved angular size of about 400 {\it milli-arcseconds} (50 AU), and 
does not show any inner structure inside of this diameter. 
The H$^{13}$CN, HC$^{15}$N, and CH$_{3}$OH
line emission regions are about twice as large as the continuum emission  and 
reveal a pronounced inner depression or "hole" with a size 
comparable to that estimated for the submillimeter continuum. We suggest that the
presence of this inner depression and the fact that we do not see inner
structure (or a flat structure) in the continuum is produced by very optically thick
dust located in the innermost parts of IRAS 16293-2422B.  
All three lines also show pronounced inverse P-Cygni profiles
with infall and dispersion velocities larger than those recently
reported from observations at lower frequencies, suggesting that we are detecting faster, 
and more turbulent gas located closer to the central object.  Finally, we report a small
east-west velocity gradient in IRAS 16293-2422B that suggests 
that its disk plane is likely located very close to the plane of the sky.
\end{abstract}


\section{Introduction}

Located at a distance of 120 pc \citep{Lau2008} in the $\rho$
Ophiuchi star forming region, IRAS 16293$-$2422B, is a well-studied
low-mass very young star. Together with its close-by companion
separated by only 600 AU (IRAS 16293$-$2422A), the entire region (IRAS
16293$-$2422) has a bolometric luminosity of 25 L$_\odot$, and is
embedded in a 2 M$_\odot$ envelope of size $\sim$ 2000 AU \citep{Cor2004}. 
Both sources show a very rich and complex chemistry, with
hot-core-like (hot-corino) properties at scales of $\sim$ 100 AU and
temperatures of about 100 K \citep{Cec1998, Caz2003, Bot2004, Cha2005, Cau2011}.

Sensitive and high angular resolution observations at 7 mm revealed
a compact, possibly isolated disk associated with IRAS 16293$-$2422B
with a Gaussian half-power radius of only 8 AU \citep{Rod2005}. 
However, the search for an outflow associated
with this source has been a hard task.  \citet{Jor2011}
using submillimeter (SMA) observations of IRAS 16293$-$2422 with
a relatively high angular resolution resolved the A and B components,
and did not find any strong indications for high velocity gas toward B. \citet{Yeh2008} 
reported the detection of a compact blue-shifted CO
structure to the south-east of source B, and mentioned that it might
correspond to a compact outflow ejected from source B. 
Loinard et al. (2012) using  ALMA observations revealed
indeed that the source B is driving a south-east blueshifted compact
outflow. However, the flow has peculiar
properties: it is highly asymmetric, bubble-like, fairly slow (10 km
s$^{-1}$), and lacking of a jet-like feature along its symmetry
axis. In addition, its dynamical age is only about 200 years.

One of the first evidences of the detection of infall motions associated
with IRAS 16293$-$2422B came from \cite{Cha2005} using 
Submillimeter Array (SMA) observations at 300 GHz. 
More recently, \citet{Pin2012} using ALMA Science Verification
observations with high-spectral resolution studied the gas kinematics
with detail in IRAS 16293$-$2422B at 220 GHz, and reported clear
inverse P-Cygni profiles toward this source in their three brightest lines 
 and derived from a simple two-layer model an infall rate of 4.5 $\times$ 10$^{-5}$
M$_\odot$ yr$^{-1}$, which is a typical value for low-mass protostars. 

In this {\it Letter}, we report $\sim$0.2 arcsec resolution 690 GHz
observations obtained with the Atacama Large Millimeter/Submillimeter
Array (ALMA) from the object IRAS 16293-2422B. The continuum
observations reveal a very compact source with a deconvolved angular 
size of about 400 {\it milli-arcseconds} or a spatial size of about 50 AU, while
the line emission shows a clear pronounced inner depression or "hole"
in the middle of IRAS 16293$-$2422B.  All the three lines mapped in
this study show pronounced inverse P-Cygni profiles.

\begin{table*}[ht]
\scriptsize 
\caption{Observational and physical parameters of the submillimeter lines}
\begin{center}
\begin{tabular}{lccccccc}
\hline \hline & Rest frequency$^a$  & E$_{lower}$ & Range of Velocities & Linewidth & LSR 
Velocity$^b$ & Line Peak flux\\ 
Lines & [GHz] & [K] & [km s$^{-1}$] & [km s$^{-1}$] & [km s$^{-1}$] &  Jy Beam$^{-1}$\\  \hline 
H$^{13}$CN [J= 8$-$7] $\nu=0$                &   690.55207 & 80.6   & $-$1,$+$7 &  4.0 & $-$3.0 & 1.00\\ 
HC$^{15}$N [J= 8$-$7]  $\nu=0$               &   688.27379 & 80.3   & $-$1,$+$7 &  3.8 & $-$3.0 & 1.00 \\ 
CH$_3$OH [9( 3, 6)- 8( 2, 7)]  $\nu_t=0$ &   687.22456 & 84.3   & $-$1,$+$7 &   3.8 & $-$3.0 & 1.02\\
 \hline 
  \hline 
\end{tabular}
\tablenotetext{a}{The rest frequencies were obtained from the Splatalogue: http://splatalogue.net}
\end{center}
\end{table*}

\section{Observations}

The observations were made with fifteen antennas of ALMA on April
2012, during the ALMA science verification data program.  The array at
that point only included antennas with diameters of 12 meters.  The
105 independent baselines ranged in projected length from 26 to 403 m.  
The observations were made in mosaicing mode using a
half-power point spacing between field centers and thus covering both
sources IRAS 16293$-$2422A and B. However, in this study we will focus
only on the molecular and continuum emission arising from IRAS
16293$-$2422B. The primary beam of ALMA at 690 GHz has a FWHM of 
$\sim 8$ arcsec.

The ALMA digital correlator was configured in 4 spectral windows of
1875 MHz and 3840 channels each.  This provides a channel width
of 0.488 MHz ($\sim$ 0.2 km s$^{-1}$), but the spectral
resolution is a factor of two lower (0.4 km/s) due to online Hanning
smoothing.

Observations of Juno provided the absolute scale for the flux density
calibration while observations of the quasars J1625$-$254 and
NRAO530 (with flux densities of 0.4 and 0.6 Jy, respectively) provided
the gain phase calibration. The quasars 3C279 and J1924-292 were used
for the bandpass calibration.

The data were calibrated, imaged, and analyzed using the Common
Astronomy Software Applications (CASA).  To analyze the data, we also
used the KARMA software \citep{Goo1996}.  The resulting r.m.s.\ noise
for the line images was about 50 mJy beam$^{-1}$  in a velocity width of 0.4 km s$^{-1}$ 
and 20 mJy beam$^{-1}$ for the continuum emission at an angular
resolution of $0\rlap.{''}31$ $\times$ $0\rlap.{''}18$ with a P.A. =
$-69.3^\circ$. We used a robust parameter of 0.5 in the CLEAN task. 
The spectra and the physical parameters
of the observed lines are shown in Figure 1 and Table 1, respectively.  
Many more spectral lines from different molecular species were 
found across the entire spectral bandwidth, however, this study will 
concentrate on the analysis of the continuum emission and the lines presented in
Table 1 that are associated with IRAS 16293$-$2422B. These selected 
lines show a good contrast between the absorption and emission 
features as compared with the rest.  We give the line peak emission
of every line in Table 1.

\section{Results and Discussion}

\subsection{0.45 mm continuum emission}

In Figure 2, we show color and contour maps of the line and continuum
emission as mapped by ALMA from IRAS 16293$-$2422B at these
wavelengths. In this Figure, we have overlaid the resulting continuum
map with the integrated intensity (moment zero) maps of the spectral
molecular lines. It is clearly observed in all lines, that the molecular emission 
surrounds the continuum emission and has a strong central depression or "hole" in
the middle.  

The peak of the continuum shows a small
offset to the west with respect to the central position of the
"hole". This small deviation might be explained by opacity effects of
the dust emission at these wavelengths.  However, this shift effect is also 
observed at longer wavelengths by \citet{Rod2005}.

\begin{figure}[ht]
\begin{center}
\includegraphics[scale=0.39]{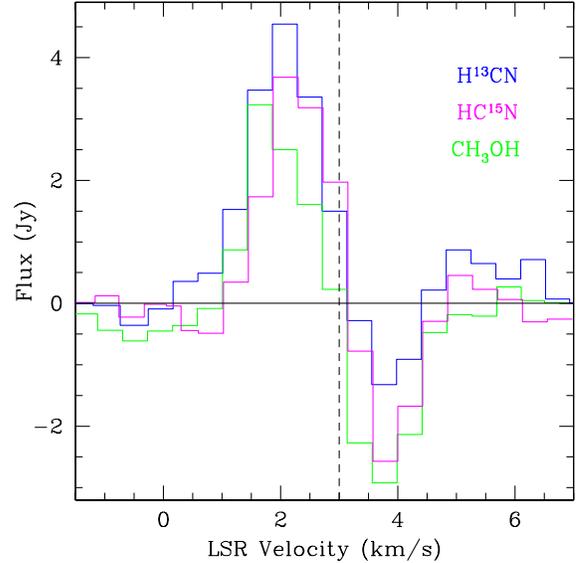}
\caption{\scriptsize H$^{13}$CN (blue), HC$^{15}$N (magenta), and 
CH$_{3}$OH (green) summed spectra from IRAS 16293$-$2422B. 
The black dashed line marks the systemic LSR velocity of 
IRAS 16293$-$2422B (V$_{LSR}$ $\sim$ 3 km s$^{-1}$). }
\label{fig1}
\end{center}
\end{figure}

The dust compact source has a deconvolved size of about 400 $\pm$ 55 {\it milli-arcseconds}
or a spatial size of 50 AU at the distance of IRAS 16293$-$2422. This
size is quite large (a factor of about six) compared to that found at 7
mm by \citet{Rod2005}. This difference in apparent angular sizes is most probably the result 
of the increasing optical depth with frequency of the dust. 
Moreover, inside of the 50 AU
diameter, the source B does not show any more inner structure even
when our beam size is about half of the source's size at these
wavelengths. This suggests that we are seeing very
optically thick dust emission at these wavelengths.

\begin{figure}
\begin{center}
\includegraphics[scale=0.39]{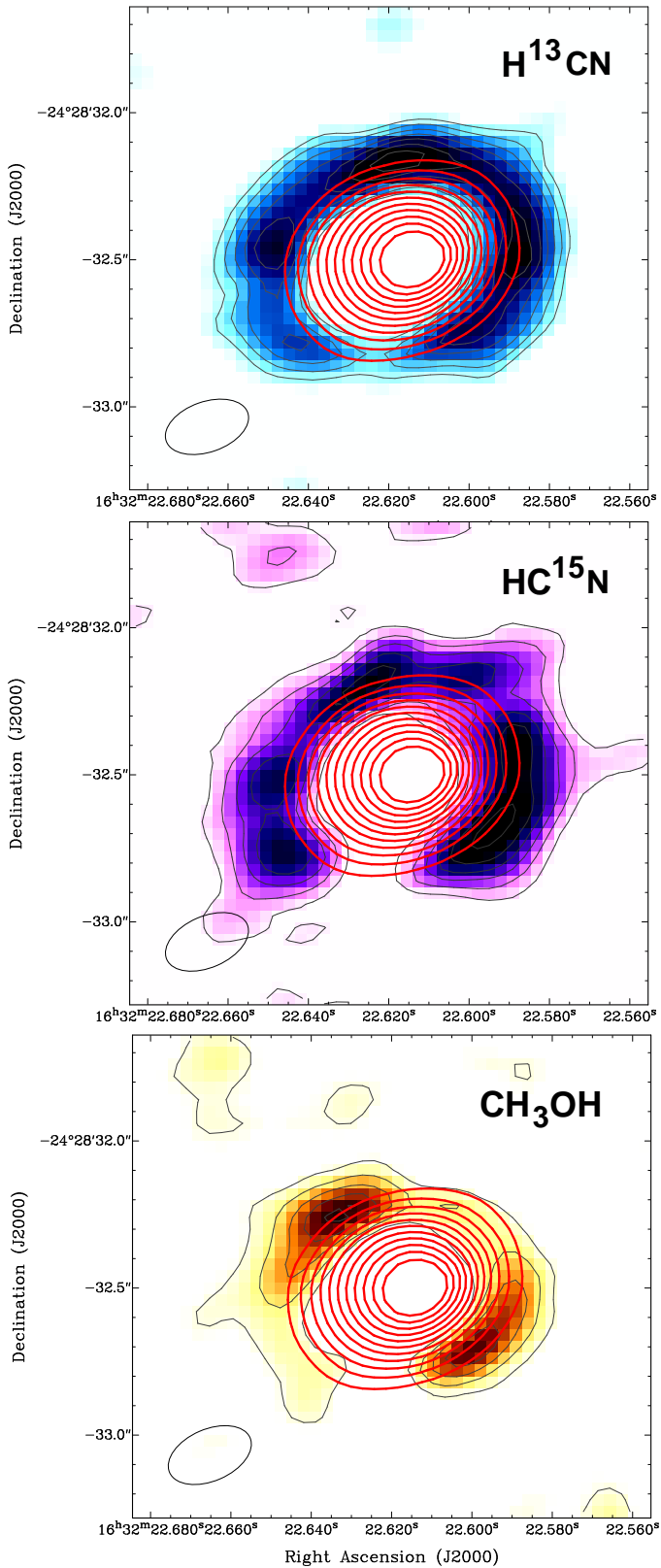}
\caption{\scriptsize H$^{13}$CN (blue), HC$^{15}$N (magenta), and
CH$_{3}$OH (yellow) color scale moment 0 maps from IRAS 16293$-$2422B
overlaid in grey contours with its own line emission and in red
contours with the 0.45 mm continuum emission. The red contours are
from 15\% to 90\% with steps of 7\% of the peak of the line emission;
the peak of the 0.45 mm continuum emission is 3.2 Jy Beam$^{-1}$. For
the line emission the grey contours are from 15\% to 90\% with steps
of 5\% of the peak of the integrated line emission. The synthesized beam of the
continuum image is shown in the bottom left corner of each image. }
\label{fig2}
\end{center}
\end{figure}

The flux density of IRAS 16293$-$2422B at these wavelengths is 12.5
$\pm$ 0.5 Jy and has a peak flux of 3.2 $\pm$ 0.2 Jy Beam$^{-1}$.
Using the full Planck equation, we can obtain the brightness temperature: 

$$
T_b = \frac{h \nu}{k  \ln \Bigg( \frac{2 h \nu^3 \Omega}{S_\nu c^2} + 1 \Bigg)},
$$

where c is the speed of light, S$_\nu$ is the flux density, $\nu$ is the frequency,
$k$ is Boltzmann constant, and $\Omega$ is the solid angle. Following this relation, and using a
Gaussian beam, we estimated a brigthness temperature at these wavelengths 
for IRAS16293$-$2422B of 160 K. 

With these flux values one can estimate a lower limit for the mass of
the disk.  Assuming that the dust is optically thin and isothermal,
the dust mass (M$_d$) will be directly proportional to the flux density (S$_\nu$) as:

$$
M_d=\frac{d^2 S_\nu}{\kappa_\nu B_\nu(T_d)},
$$    

\begin{figure}[ht]
\begin{center}
\includegraphics[scale=0.5]{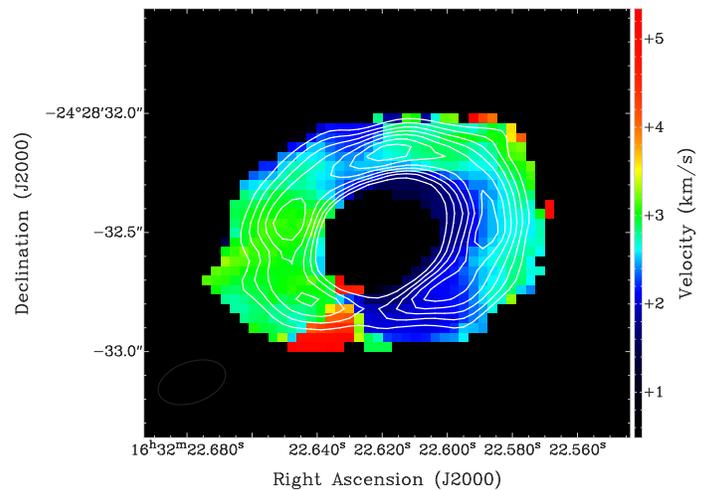}
\caption{\scriptsize Integrated intensity of the weighted velocity
(moment 1) color map of the H$^{13}$CN emission from the source IRAS
16293$-$2422B overlaid in contours with the integrated intensity
contour map (white).  The white contours are from 15\% to 90\% with
steps of 5\% of the peak of the integrated line emission; the peak of the integrated line
emission is 4.1 $\times$ 10$^3$ mJy Beam$^{-1}$ km s$^{-1}$.  The
color-scale bar on the right indicates the LSR velocities in km
s$^{-1}$.  
}
\label{fig1}
\end{center}
\end{figure}

\noindent
where the $d$ is the distance to the object, $\kappa_\nu$ the dust
mass opacity, and B$_\nu(T_d)$ the Planck function for the dust
temperature T$_d$.  Assuming a dust mass opacity ($\kappa_\nu$) of 2.2
cm$^2$ g$^{-1}$, obtained extrapolating at these wavelenghts
the value obtained by \citet{Oss1994} 
for coagulated dust particles with no ice mantles, and at a density of 10$^8$ cm$^{-3}$. 
Assuming also an opacity power-law index $\beta$ = 0.6 \citep{Rod2005},
as well as a characteristic dust temperature (T$_d$) of 160 K, we
estimated a lower limit for the mass for the disk of about 0.03 M$_\odot$. 
Please note that the level of uncertainty in the mass lower limit 
is a factor of five, given the range of 0.435 mm opacities in Table 1
of \citet{Oss1994}. 

\subsection{Molecular line emission}

In Figure 1 and 2, as mentioned earlier, we present the integrated intensity maps (moment 0) 
of the molecular emission reported on this
work. The spectrum from all lines were obtained averaging in an area (box)
similar to the size of the molecular ring like structure ($\sim$ 1.0$''$). 
The spectrum of all lines is found to be well centered at an LSR
velocity of $+$3 km s$^{-1}$, which is approximately the systemic velocity for this
source \citep{Pin2012, Jor2011}.  All three lines
also show marked inverse P-Cygni profiles but with that of the CH$_3$OH showing
the more pronounced absorption feature. This is probably due to this line
frequently being more optically thick.  The H$^{13}$CN and HC$^{15}$N
show line profiles very similar with the emission components being
stronger compared with the CH$_3$OH spectra, see Figure 1.  This
latter line mostly shows two faint condensations surrounding the
continuum source and does not present a marked ring like structure as compared 
with the rest of the lines.

The morphology of the line emission in general forms a well defined ring like structure around the
continuum. However, the ring like structure is not completely closed, there is a small
cavity towards its southeast. This cavity is  probably created by the
southeast monopolar outflow reported by Loinard et al. (2012). 
The molecular ring like structure has a diameter of 850 $\pm$ 50 {\it milli-arcseconds} 
and an inner diameter of 300 $\pm$ 50 {\it milli-arcseconds}, which is comparable to
the deconvolved size of the 0.45 mm continuum source (about 400 {\it milli-arcseconds}).

\begin{figure}[ht]
\begin{center}
\includegraphics[scale=0.46]{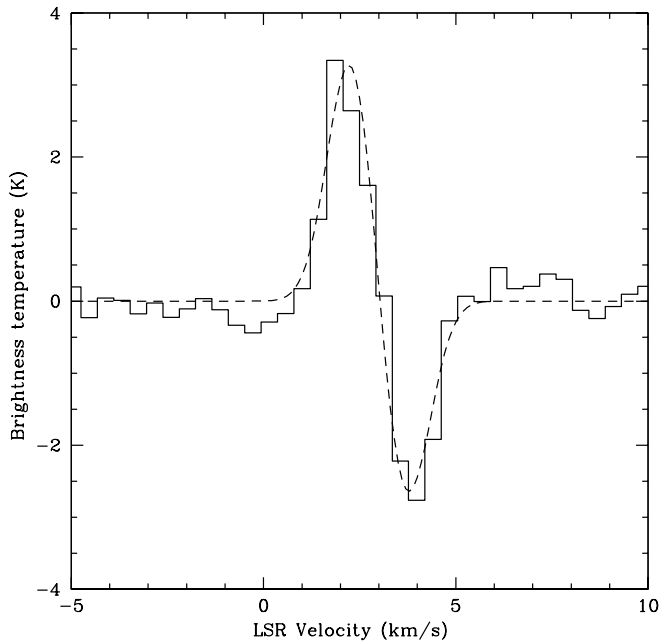}
\caption{\scriptsize  Observed (solid line) and 
modeled (dashed line) spectra of the CH$_{3}$OH. }
\label{fig1}
\end{center}
\end{figure}


In Figure 3, we show the integrated intensity of the weighted velocity
(moment 1) color map of the H$^{13}$CN.  This figure reveals a clear
east-west velocity gradient of approximately 1.3 km s$^{-1}$ over 60 AU.  
This small gradient might suggest that the orientation of the disk plane is very
close to the plane of the sky.  If we assume that the rotation is
Keplerian and attribute it to a disk in rotation then the dynamical
mass associated to this velocity gradient corresponds to only  0.1 M$_\odot$, 
which is indeed very small and is comparable with the dust disk mass.  
However, if we correct by the inclination angle dividing this mass 
by the $\sin \theta$ with $\theta$ say 5$^\circ$, we obtain a value 
of 1.2 M$_\odot$, a more reasonable value for the central object and 
the disk associated with a solar type young star,
see \citet{Cor2004}. We therefore conclude that the disk plane of IRAS16293$-$2422B 
must be located almost on the plane of the sky, as already suggested 
by \citet{Rod2005}.

\subsection{Modeling}

In Figure 4, we show the result of the modeled spectra
that we made in this study.
We fitted the spectral profile using a modified
two-layer model (Myers et al. 1996; Di Francesco et
al 2001; Kristensen et al. 2012), as
described by Pineda et al. (2012).
This model consists of two layers of gas, front and rear, 
that are infalling
towards the central source with an infall velocity, velocity
dispersion, excitation temperature, and opacity at the center
of the line of $V_{in}$, $\sigma_v$, 
$T_x$, and $\tau_0$, respectively.
In between the two layers there is an optically thick
continuum source
emitting as a blackbody of temperature $T_c$, and filling a fraction
of the beam, $\Phi$.
The background temperature that illuminates the rear layer
is taken to be the cosmic background, $T_f$ = 3 K.

The brightness temperature of the optically thick continuum source
is taken to be such that it matches the peak continuum
flux density of the image, $T_c$ = 160 K. The adopted filling
fraction of the continuum source, $\Phi$ =0.37 is consistent with
the ratio of solid angles between the region in absorption
to the region in emission. From the fits of Pineda et al. (2012)
to lines at lower frequencies (220 GHz) they estimate
$T_x$ = 40 K. The lines sampled by us are probably
originating in gas closer to the star (see below). Assuming that
the excitation temperature of the molecules decreases as the square root of 
the distance and that the gas sampled by Pineda et al. (2012)
is 50\% more distant than that sampled by us, we adopt 
$T_x$ = 50 K.  The fitting was minimized with a grid search,
obtaining the following parameters: $V_{in}$ = 0.7 km s$^{-1}$,
$\sigma_v$ = 0.6 km s$^{-1}$, and $\tau_0$ =0.17. Finally, the
systemic velocity obtained from the fit was $V_{LSR}$ = 3.0 km s$^{-1}$.

Several of the parameters are quite similar to those derived by
Pineda et al. (2012) from lines at 220 GHz. However, others are
not and we discuss them here. First, the brightness temperature of the optically thick 
continuum source is 20 K in the case of Pineda et al. (2012) and
160 K in this paper. This is as expected since the optical depth of
dust increases sharply with frequency. The large brightness temperature
derived by us and consistent with our profile modeling implies
that at 690 GHz we are observing a truly optically-thick disk since
the brightness temperature is comparable with the thermodynamic
temperatures expected in the inner parts of a YSO accretion disk.
The optical depth used by us is about one half of that used by Pineda
et al. (2012). Finally, the infall velocity and velocity
dispersion required by our modeling (0.7 and 0.6 km s$^{-1}$, respectively) 
are larger than those used by Pineda et al. (about 0.5 and 0.3 km s$^{-1}$, respectively), 
implying that we may be detecting faster, more turbulent gas located
closer to the central object. 
This is consistent with the standard picture of infall, 
where higher velocities occur at smaller radii. The velocities reported here are
supersoinic as those reported in Pineda et al. (2012).









\section{Summary}

In this paper, we have reported line and continuum observations 
obtained with ALMA at 690 GHz with a very high angular resolution 
($\sim$ 0.2 arcsec) of IRAS 16293-2422B. The main conclusions are as follows:

\begin{itemize}

\item The 0.45 mm continuum emission revealed a very compact object with a
deconvolved angular size of about 400 {\it milli-arcseconds} that is associated with IRAS 16293-2422B.
This size is very large compared to the one reported at 7 mm (about 8 AU)
and does not show any structure inside of this diameter (or a flat structure). 

\item The H$^{13}$CN, HC$^{15}$N, and CH$_{3}$OH  images revealed a
pronounced inner depression or "hole" with a size comparable to that
estimated for the submillimeter continuum. 

\item We suggest that the presence of this inner depression with an angular size comparable with 
that of the continuum source and the fact that we do not see inner structure in the continuum 
is produced by very optically thick dust located in the innermost parts of IRAS 16293-2422B.  

\item All three lines also show inverse P-Cygni profiles
with infall and dispersion velocities larger than those recently
reported at smaller wavelengths, suggesting that we are revealing faster, and more turbulent
gas located closer to the central object.  

\item We report a small east-west velocity gradient in IRAS 16293-2422B observed
 in all lines that suggests that the disk plane of this object is likely located very close to the plane 
of the sky.
\end{itemize}

\acknowledgments
L.A.Z, L. L. and L. F. R. acknowledge the financial support from
DGAPA, UNAM, and CONACyT, M\'exico.  L. L. is indebted to the
Alexander von Humboldt Stiftung for financial support. 
This paper makes use of the following ALMA data: ADS/JAO.ALMA\#2011.0.00007.SV. 
ALMA is a partnership of ESO (representing its member
states), NSF (USA) and NINS (Japan), together with NRC (Canada) and
NSC and ASIAA (Taiwan), in cooperation with the Republic of Chile. The
Joint ALMA Observatory is operated by ESO, AUI/NRAO and NAOJ.


\begin{thebibliography}{}

\bibitem[Bottinelli et al.(2004)]{Bot2004} Bottinelli, S., 
Ceccarelli, C., Neri, R., et al.\ 2004, \apjl, 617, L69 

\bibitem[Caux et al.(2011)]{Cau2011} Caux, E., Kahane, C., Castets,
A., et al.\ 2011, \aap, 532, A23

\bibitem[Cazaux et al.(2003)]{Caz2003} Cazaux, S., Tielens,
A.~G.~G.~M., Ceccarelli, C., et al.\ 2003, \apjl, 593, L51

\bibitem[Ceccarelli et al.(1998)]{Cec1998} Ceccarelli, C., Castets,
A., Loinard, L., Caux, E., \& Tielens, A.~G.~G.~M.\ 1998, \aap, 338,
L43

\bibitem[Chandler et al.(2005)]{Cha2005} Chandler, C.~J., Brogan,
C.~L., Shirley, Y.~L., \& Loinard, L.\ 2005, \apj, 632, 371

\bibitem[Correia et al.(2004)]{Cor2004} Correia, J.~C., Griffin, M.,
\& Saraceno, P.\ 2004, \aap, 418, 607

\bibitem[Di Francesco et al.(2001)]{Di2001} Di Francesco, J., Myers,
P.~C., Wilner, D.~J., Ohashi, N., \& Mardones, D.\ 2001, \apj, 562,
770

\bibitem[Gooch(1996)]{Goo1996} Gooch, R.\ 1996, Astronomical Data
Analysis Software and Systems V, 101, 80


\bibitem[J{\o}rgensen et al.(2011)]{Jor2011} J{\o}rgensen, J.~K.,
Bourke, T.~L., Nguyen Luong, Q., \& Takakuwa, S.\ 2011, \aap, 534,
A100

\bibitem[Kristensen et al.(2012)]{Kri2012} Kristensen, L.~E., van
Dishoeck, E.~F., Bergin, E.~A., et al.\ 2012, \aap, 542, A8

\bibitem[Loinard et al.(2008)]{Lau2008} Loinard, L., Torres, R.~M.,
Mioduszewski, A.~J., \& Rodr{\'{\i}}guez, L.~F.\ 2008, \apjl, 675, L29

\bibitem[Loinard et al.(2012)]{Lau2012} Loinard, L., Zapata, 
L.~A., Rodr{\'{\i}}guez, L.~F., et al.\ 2012, \mnras, L35 

\bibitem[Myers et al.(1996)]{Mye1996} Myers, P.~C., Mardones, D.,
Tafalla, M., Williams, J.~P., \& Wilner, D.~J.\ 1996, \apjl, 465, L133

\bibitem[Ossenkopf \& Henning(1994)]{Oss1994} Ossenkopf, V., \&
Henning, T.\ 1994, \aap, 291, 943

\bibitem[Pineda et al.(2012)]{Pin2012} Pineda, J.~E., Maury, A.~J.,
Fuller, G.~A., et al.\ 2012, \aap, 544, L7

\bibitem[Rodr{\'{\i}}guez et al.(2005)]{Rod2005} Rodr{\'{\i}}guez,
L.~F., Loinard, L., D'Alessio, P., Wilner, D.~J., \& Ho, P.~T.~P.\
2005, \apjl, 621, L133


\bibitem[Yeh et al.(2008)]{Yeh2008} Yeh, S.~C.~C., Hirano, N., Bourke,
T.~L., et al.\ 2008, \apj, 675, 454



\end{thebibliography}
\end{document}